\documentclass[]{imsart}

\RequirePackage{amsthm,amsmath,amsfonts,amssymb}
\RequirePackage[authoryear]{natbib}
\RequirePackage[colorlinks,citecolor=blue,urlcolor=blue]{hyperref}
\RequirePackage{graphicx}
\RequirePackage{xcolor}

\begin{document}

\begin{frontmatter}

\title{Bayesian Modeling of Effective and Functional Brain Connectivity
using Hierarchical Vector Autoregressions}

\runtitle{Bayesian VAR Hierarchical Model for Brain Connectivity}

\begin{aug}

\author[A]{\fnms{Bertil} \snm{Wegmann}\ead[label=e1]{bertil.wegmann@liu.se}},
\author[B]{\fnms{Anders} \snm{Lundquist}\ead[label=e2]{anders.lundquist@umu.se}}
\author[C]{\fnms{Anders} \snm{Eklund}\ead[label=e3]{anders.eklund@liu.se}}
\and
\author[D]{\fnms{Mattias} \snm{Villani}\ead[label=e4]{mattias.villani@gmail.com}}

\address[A]{Div. of Statistics and Machine Learning, Dept. of Computer and Information Science, Linköping University, \printead{e1}}
\address[B]{Div. of Statistics, Umeå School of Business and Economics (USBE), Umeå University, \printead{e2}}
\address[C]{Div. of Medical Informatics, Dept. of Biomedical Engineering, Linköping University, \printead{e3}}
\address[D]{Dept. of Statistics, Stockholm University, \printead{e4}}
\end{aug}

\begin{abstract}
Analysis of brain connectivity is important for understanding how
information is processed by the brain. We propose a novel Bayesian
vector autoregression (VAR) hierarchical model for analyzing brain
connectivity in a resting-state fMRI data set with autism spectrum
disorder (ASD) patients and healthy controls. Our approach models
functional and effective connectivity simultaneously, which is new
in the VAR literature for brain connectivity, and allows for both
group- and single-subject inference as well as group comparisons.
We combine analytical marginalization with Hamiltonian Monte Carlo
(HMC) to obtain highly efficient posterior sampling. The results from
more simplified covariance settings are, in general, overly optimistic
about functional connectivity between regions compared to our results.
In addition, our modeling of heterogeneous subject-specific covariance
matrices is shown to give smaller differences in effective connectivity 
compared to models with a common covariance matrix to all subjects.
\end{abstract}


\end{frontmatter}

\section{Introduction}

The use of functional magnetic resonance imaging (fMRI) to investigate
brain connectivity dates back to seminal papers from the mid-1990:s
(\citealt{Friston1994a}, \citealt{Biswal1995}), and during the last
decade the interest has increased dramatically (\citealt{Solo2018}).
Brain connectivity is an important tool in understanding how information
is processed by the brain, with applications in both non-clinical
and clinical settings. On the clinical side, connectivity is frequently
investigated in association with different kinds of neuropsychological
conditions, e.g. schizophrenia (\citealt{Lynall2010}), ADHD (\citealt{Konrad2010}),
epilepsy (\citealt{Morgan2015}) and autism spectrum disorder (ASD)
(\citealt{Easson2019}). The most common approach for studying brain
connectivity using fMRI is so-called resting state fMRI (rs-fMRI),
and the analyses are mainly divided into functional and effective
connectivity. Functional connectivity pertains to the investigation
of undirected associations between brain regions, whereas effective
connectivity refers to directed associations (\citealt{Friston1994a,Friston2011}).
Our work targets both functional and effective connectivity, which
is new in the literature, and the possibility to perform both group-level
inference, which is most common and important in practice, as well
as subject-specific inference.

In an rs-fMRI scan, the subject is typically scanned for around ten
minutes or less, without specific instructions or tasks to complete.
The brain is divided into three-dimensional pixels, called voxels,
with a side of about 3-4 mm. Within each voxel, the blood oxygenation
level depend (BOLD) signal is measured, which is a proxy measure for
neural activity. The BOLD signal in every voxel is usually sampled
at about 0.5-2 seconds intervals; for a rigorous and detailed introduction
to fMRI, see \citet{buxton2009introduction}. The data obtained are
therefore voxelwise time series, where the time series length is usually
200-1000 observations. The number of voxels tend to be in the hundreds
of thousands, and some kind of dimension reduction is therefore typically
used to divide the brain into at most a few hundred functional regions
using a brain atlas, see e.g. \citet{Power2011} or \citet{Glasser2016}.
A single time series for each region is then constructed by e.g. averaging
the time series for all voxels within the region. Our work assumes
the rs-fMRI data has been preprocessed to yield such regionwise time
series.

The overarching statistical problem in functional and effective connectivity
analysis is to estimate suitable measures of association between brain
regions. The perhaps simplest solution, which is still widely used
in practice, is to calculate the pairwise Pearson correlation, or
partial correlation, between every pair of regions. These approaches
have the obvious drawback of completely disregarding autoregressive
dependencies, both within and between regions. \citet{afyouni2019effective}
point out that the standard error of the sample correlation coefficient
becomes biased from autocorrelation, which implies that the commonly
used Fisher transformation can not stabilise the variance.

Attention has therefore turned to time series modeling, e.g. wavelet
expansions (\citealt{Zhang2014}) and vector autoregressive (VAR)
models either in the time domain (\citealt{Chiang2017}, \citealt{Goebel2003}),
or in the frequency domain (\citealt{Cassidy2015,Cassidy2018}). Dynamic
Causal Models (DCM, \citealt{Friston2003}), a state-space model with
ambitious neurophysiological modeling have been extended from task
fMRI data to analyze connectivity from resting-state data (\citealt{Friston2014}).

We propose a Bayesian hierarchical VAR model for both effective and
functional connectivity that accounts for autoregressive dependencies
within and between brain regions. The hierarchical setting allows
for comparison of group-level inference in a straightforward manner.
Our results differ substantially from other comparable VAR models
in the brain connectivity literature. In general, the results from
more simplified covariance settings overestimate functional connectivity
between regions compared to our results. We also observe smaller differences
in EC, which we suspect are mainly due to heterogeneous subject-specific
covariance matrices, which other approaches with a common covariance
matrix for all subjects can not account for. We fit these more complex
and computationally very demanding models by integrating out the subject-specific
parameters to obtain posterior inference on the group-level parameters
using highly efficient Hamiltonian Monte Carlo (HMC) sampling. The
article is organized as follows. In Section 2, we define our proposed
model with subject-specific covariance matrices, and models with more
simplified covariance settings (\citet{Chiang2017} and \citet{Gorrostieta2012a,Gorrostieta2013}).
Our Bayesian setting is described and posterior inference is derived
in Section 3. Group-level inference on a real rs-fMRI data set with
controls and individuals diagnosed with ASD is presented in Section
4. Concluding remarks are given in Section 5.

\section{A Bayesian VAR hierarchical modeling for Brain Connectivity\label{sec:Models}}

This section describes a Bayesian vector autoregressive (VAR) hierarchical
model for brain connectivity applied to a group of subjects. The model
allows for subject-specific VAR parameters centered around a group-level
VAR. We also discuss two special cases of our model which have been
used for effective connectivity by \citet{Chiang2017} and \citet{Gorrostieta2012a,Gorrostieta2013}.
Section \ref{sec:Bayesian-Inference} proposes an efficient posterior
sampling algorithm for the model that combines analytical marginalization
with HMC.

\subsection{The hierarchical VAR model}

Let $Y_{rst}$ be the fMRI BOLD signal for subject $s$ in region
$r$ at time $t$, where $r=1,\dots,R,$ $s=1,\dots,S,$ and $t=1,\dots,T$.
The Bayesian VAR (BVAR) model of order $L$ can be defined for 
\[
Y_{s,t}=\left[\begin{array}{c}
Y_{1st}\\
\vdots\\
Y_{Rst}
\end{array}\right],B_{ls}=\left[\begin{array}{ccc}
B_{11ls} & \cdots & B_{1Rls}\\
\vdots & \ddots & \vdots\\
B_{R1ls} & \cdots & B_{RRls}
\end{array}\right],B_{s}=\left(B_{1s},\dots,B_{Ls}\right)
\]
as
\begin{gather}
Y_{s,t}=\sum_{l=1}^{L}B_{ls}Y_{s,t-l}+\epsilon_{st},\;\epsilon_{st}\sim N_{R}\left(0,\Sigma_{s}\right)\nonumber \\
\textrm{vec}\left(B_{s}\right)|\Sigma_{s}\sim N_{LR^{2}}\left(\textrm{vec}\left(B\right),\Sigma_{s}\otimes P_{s}^{-1}\right)\nonumber \\
\Sigma_{s}\sim IW\left(\nu\Sigma,\nu\right),\;f\left(\nu\right)\propto1,\,\nu>R+1\nonumber \\
\textrm{vec}\left(B\right)|\Sigma\sim N\left(B_{0},\Sigma\otimes P_{0}^{-1}\right)\label{eq:hierVAR}\\
\Sigma\sim IW\left(\nu_{0}\Psi_{0},\nu_{0}\right),\nonumber 
\end{gather}
where $N_{k}\left(\right)$ and $IW\left(\right)$ denote the $k$-dimensional
multivariate normal and inverse Wishart distributions, respectively.
Thus, the user needs to specify prior precision matrices $P_{0}$,
$\Psi_{0}$, $P_{s}$ for $s=1,\ldots,S$\textasciiacute , and the
degrees of freedom $\nu_{0}>R+1$. We elaborate on our chosen prior
specification in Section \ref{subsec:PriorSpec}. We label the general
hierarchical VAR in \eqref{eq:hierVAR} as Model 1.

The global parameters $B$ and $\Sigma$ in \eqref{eq:hierVAR} are
of main interest, in particular comparing the connectivity implied
by $B$ and $\Sigma$ between groups of subjects; see the application
in Section \ref{sec:Results} where a group of subjects diagnosed
with autism spectrum disorder (ASD) are compared to healthy controls.

We also consider the following two submodels of Model 1. First, Model
$2$ assumes a common covariance matrix for all subjects, i.e. $\Sigma_{s}=\Sigma$
for all $s$. This model is clearly nested in Model $1$, since
\[
\left(\Sigma_{s}\right)_{\nu}\overset{d}{\longrightarrow}\Sigma\text{ as }\nu\longrightarrow\infty.
\]
Model 3 makes the additional simplification also made by \citet{Chiang2017}
and \citet{Gorrostieta2012a,Gorrostieta2013}; that the common covariance
matrix $\Sigma$ is diagonal with prior independent diagonal elements
following a conjugate inverse gamma ($IG$) distribution (\citealt{Press2005}),
\[
\Sigma_{rr}\sim IG\left(\frac{\nu_{0}-2R}{2},\frac{\nu_{0}\Psi_{0rr}}{2}\right).
\]
Hence, Model 3 does not allow estimation of functional connectivity
from non-zero off-diagonal elements of $\Sigma$.

\section{Bayesian Inference\label{sec:Bayesian-Inference}}

The Bayesian approach updates a prior distribution for all model parameters
with observed data through the likelihood function to a posterior
distribution
\begin{equation}
p\left(B_{1:S},\Sigma_{1:S},B,\Sigma|Y\right)\propto p\left(Y|B_{1:S},\Sigma_{1:S},B,\Sigma\right)p\left(B_{1:S},\Sigma_{1:S},B,\Sigma\right),\label{eq:posteriorAll}
\end{equation}
where $B_{1:S}=\{B_{1},\ldots,B_{S}\}$, and $\Sigma_{1:S}$ is defined
analogously. The object of main interest is the marginal posterior
of the group-level parameters

\begin{equation}
p\left(B,\Sigma|Y\right)\propto p\left(Y|B,\Sigma\right)p\left(B,\Sigma\right),\label{eq:marginalposterior}
\end{equation}
which is obtained by integrating out the subject-specific parameters
from \eqref{eq:posteriorAll}. The marginal posterior distribution
in Model 2, i.e. the model with common $\Sigma$ for all subjects,
can be derived in closed form. However, the posterior for the general
model where all parameters are subject-specific is not tractable.
We propose an efficient HMC algorithm to sample from the marginal
posterior in \eqref{eq:marginalposterior}. The analytical result
for Model 2 is also exploited for determining the prior hyperparameters
in Model 1.

\subsection{Posterior inference when $\Sigma$ is common to all subjects\label{subsec:PosteriorModel2and3}}

Let $X_{s}=\left(Y_{s,t-1},\dots,Y_{s,t-L}\right)$ be the set of
covariates in the BVAR model. The likelihood function of $\left(B_{s},\Sigma\right)$
for each subject $s$ is given by (see Appendix A for details)
\begin{equation}
p\left(Y_{s}|B_{s},\Sigma,X_{s}\right) = \left|2\pi\Sigma\right|^{-n/2}\exp\left(-\frac{1}{2}\textrm{tr}\Sigma^{-1}V_{s}\right)\label{eq:Likelihood}
\end{equation}
\[
\times\exp\left(-\frac{1}{2}\left(B_{s}-\hat{B_{s}}\right)^{T}\left(\Sigma^{-1}\otimes X_{s}^{T}X_{s}\right)\left(B_{s}-\hat{B_{s}}\right)\right),
\]
where $n=T-L$, $V_{s}=\left(Y_{s}-X_{s}\hat{B_{s}}\right)^{T}\left(Y_{s}-X_{s}\hat{B_{s}}\right)$
and $\hat{B_{s}}=\left(X_{s}^{T}X_{s}\right)^{-1}X_{s}^{T}Y_{s}.$
The marginal likelihood function of $\left(B,\Sigma\right)$ for all
subjects is given by 
\[
\prod_{s=1}^{S}p\left(Y_{s}|B,\Sigma,X_{s}\right)=\prod_{s=1}^{S}\int p\left(Y_{s}|B_{s},\Sigma,X_{s}\right)p\left(B_{s}|\Sigma\right)dB_{s}.
\]
Multiplying this marginal likelihood with the prior distribution of
$\left(B,\Sigma\right)$, the posterior distribution of $\left(B,\Sigma\right)$
becomes (see Appendix A for details)
\begin{equation}
p\left(B,\Sigma|Y_{s},X_{s}\right)=c_{0}c_{\kappa}\left|P_{0}\right|^{p/2}\left|\Sigma\right|^{-\left(Sn+\nu_{0}+p+1\right)/2}\exp\left(-\frac{1}{2}\textrm{tr}\Psi_{n}\Sigma^{-1}\right)\label{eq:Posterior_B_Sigma}
\end{equation}
\[
\times\exp\left(-\frac{1}{2}\textrm{tr}\Sigma^{-1}\left(B-\tilde{B}\right)^{T}\tilde{P}\left(B-\tilde{B}\right)\right),
\]
where $c_{0}$ does not depend on $B$ and $\Sigma,$ $c_{\kappa}=\prod_{s=1}^{S}c_{\kappa_{s}}=\prod_{s=1}^{S}\left(\left|P_{s}\right|^{p/2}\left|P_{s}+X_{s}^{T}X_{s}\right|^{-p/2}\right)$,
$\Psi_{n}=\nu_{0}\Psi_{0}+\sum_{s=1}^{S}\left(R_{s}+E_{s}^{T}Q_{s}^{-1}E_{s}\right)+B_{0}^{T}P_{0}B_{0}-\tilde{B}^{T}\tilde{P}\tilde{B},$\\
$\tilde{P}=P_{0}+\sum_{s=1}^{S}Q_{s}^{-1},$ $\tilde{B}=\tilde{P}^{-1}\left(P_{0}B_{0}+\sum_{s=1}^{S}Q_{s}^{-1}E_{s}\right),$\\
$R_{s}=\left(Y_{s}-X_{s}K_{1s}X_{s}^{T}Y_{s}\right)^{T}\left(Y_{s}-X_{s}K_{1s}X_{s}^{T}Y_{s}\right)+Y_{s}^{T}X_{s}K_{1s}P_{s}K_{1s}X_{s}^{T}Y_{s}-E_{s}^{T}Q_{s}^{-1}E_{s}$,
\\
$K_{1s}=\left(P_{s}+X_{s}^{T}X_{s}\right)^{-1},$ $Q_{s}=\left(P_{s}K_{1s}X_{s}^{T}X_{s}K_{1s}P_{s}+\left(I-K_{1s}P_{s}\right)^{T}P_{s}\left(I-K_{1s}P_{s}\right)\right)^{-1},$
\\
$E_{s}=Q_{s}\left(P_{s}K_{1s}X_{s}^{T}\left(Y_{s}-X_{s}K_{1s}X_{s}^{T}Y_{s}\right)+\left(I-K_{1s}P_{s}\right)^{T}P_{s}K_{1s}X_{s}^{T}Y_{s}\right)$
and $I$ is the identity matrix.

Conditional on $\Sigma$ the posterior distribution of $B$ is given
by 
\[
B|\Sigma,\boldsymbol{Y}\sim N_{p,q}\left(\tilde{B},\tilde{P}^{-1},\Sigma\right),
\]
i.e. a matrix-Normal distribution with posterior mean $\tilde{B}$
as a weighted average of the \emph{data mean} $B_{D}=\left(\sum_{s=1}^{S}Q_{s}^{-1}\right)^{-1}\sum_{s=1}^{S}Q_{s}^{-1}E_{s}$
and \emph{prior mean} $B_{0}$. Integrating out $B$, the marginal
posterior distribution of $\Sigma$ is 
\[
\Sigma|\boldsymbol{Y}\sim IW\left(\Psi_{n},\,\nu_{n}\right),
\]
where $\nu_{n}=\nu_{0}+Sn-q.$ Hence, the marginal posterior distribution
of $\Sigma$ is an Inverse-Wishart distribution with $\nu_{n}$ degrees
of freedom and scale matrix $\Psi_{n}$ in Model 2. This implies for
Model 3 that the marginal posterior distribution of each element $rr$
in the diagonal matrix $\Sigma$ follows an inverse gamma distribution
as (\citealt{Press2005})
\[
\Sigma_{rr}|\boldsymbol{Y}\sim IG\left(\frac{\nu_{n}-2R}{2},\frac{\nu_{n}\Psi_{nrr}}{2}\right).
\]

\subsection{Posterior inference for the hierarchical VAR with subject-specific
$\Sigma_{s}$}

Replacing $\Sigma$ with $\Sigma_{s}$ in Equation \eqref{eq:Likelihood}
gives the likelihood function of $\left(B_{s},\Sigma_{s}\right)$
for each subject $s$. Then, the marginal likelihood function of $\left(B,\Sigma,\nu\right)$
becomes (see Appendix B for details)

\[
p\left(Y|B,\Sigma,\nu,X\right)=\prod_{s=1}^{S}\int\int p\left(Y_{s}|B_{s},\Sigma_{s},X_{s}\right)p\left(B_{s},\Sigma_{s}\vert B,\Sigma\right)dB_{s}d\Sigma_{s}
\]
\[
=c_{1}\left|\nu\Sigma\right|^{\nu/2}\prod_{s=1}^{S}\left|\nu\Sigma+R_{s}+\left(B-E_{s}\right)^{T}Q_{s}^{-1}\left(B-E_{s}\right)\right|^{-\left(n+\nu\right)/2},
\]
where $c_{1}$ does not depend on $B$, $\Sigma,$ and $\nu$. The
posterior distribution of $\left(B,\Sigma,\nu\right)$ is intractable
and high-dimensional, so we use the HMC algorithm with hyperparameters
tuned adaptively using the No-U-Turn Sampler (NUTS) \citep{hoffman2014no}
to sample from the posterior. We implement the algorithm in the probabilistic
programming language Stan, see Appendix C for the Stan model specification.
To monitor convergence to the posterior, we run three parallel MCMC
chains until the diagnostic convergence measure $\hat{R}$ in \citet{Gelman1992}
is close to $1$.

\subsection{Prior specification\label{subsec:PriorSpec}}

Let $\textrm{max }s_{r}^{2}$ be the maximum sample variance in region
$r$ for all subjects. We choose a non-informative prior for $\Sigma$
by letting $\Psi_{0}$ in \eqref{eq:hierVAR} be a diagonal matrix
with elements $\Psi_{0rr}=\textrm{max }s_{r}^{2},\;r=1,\dots,R,$
and a low degree of freedom $\nu_{0}=R+2$. Following \citet{Litterman1986},
it is common practice in the BVAR literature to impose heavier shrinkage
on higher lag orders. To implement this effect we let $P_{0}^{-1}=\lambda D$
and $P_{s}^{-1}=\kappa_{s}D$ for each subject $s$, where the diagonal
elements of $D$ for lag $l$ are given by $\left(l^{2}\bar{s_{r}^{2}}\right)^{-1},$
where $\bar{s_{r}^{2}}$ is the mean of the subjects' sample variances
in region $r$.

The values of $\lambda$ and $\kappa_{s}$ are obtained from an empirical
Bayes approach by maximizing the analytical, tractable, marginal likelihood
of $Y$ in Model $2$. This is expected to be a good approximation
to the optimal hyperparameters for Model 1 since $\lambda$ and $\kappa_{s}$
are not related to $\Sigma$ or $\Sigma_{s}$, which is the aspect
that differs between Models 1 and 2. Integrating out $\left(B,\Sigma\right)$
from the posterior distribution in Equation \eqref{eq:Posterior_B_Sigma},
the marginal likelihood of the data $\boldsymbol{Y}$ can be written
as a function of $\lambda$ and $\kappa=\left(\kappa_{1},\dots,\kappa_{S}\right)$
as
\[
f\left(\boldsymbol{Y},\kappa,\lambda\right)=c_{2}c_{\kappa}\left|P_{0}\right|^{p/2}\left|\tilde{P}\right|^{-p/2}\left|\Psi_{n}\right|^{-\frac{1}{2}\left(Sn-q+\nu_{0}\right)},
\]
where $c_{2}$ does not depend on $\kappa$ and $\lambda$. Optimizing
this function with respect to $\lambda$ and $\kappa$, gives the
estimated values of the hyperparameters in the prior precision matrices
$P_{0}$ and $P_{s}$ of all models, respectively.

\section{Brain Connectivity in resting-state fmri data\label{sec:Results}}

In Section \ref{subsec:Data}, we describe the data used for group
comparisons between healthy controls and individuals diagnosed with
ASD. Effective and functional connectivity results are presented and
compared between the models in Section \ref{subsec:ECFCres}. In Section
\ref{subsec:ComputationTimes}, we present a brief overview of computational
time requirements for analyzing the data with different number of
time lags and number of regions considered.

\subsection{Description of data and ROI selection\label{subsec:Data}}

We use data from ABIDE\footnote{http://fcon\_1000.projects.nitrc.org/indi/abide/abide\_I.html}
\citep{di2014autism} preprocessed\footnote{http://preprocessed-connectomes-project.org/abide/index.html}
\citep{craddock2013neuro} consisting of resting state fMRI data from
539 individuals diagnosed with ASD and 573 healthy controls; we use
randomly selected subsets of 20 controls and 20 ASD patients from
the data collected at New York University. The fMRI data were collected
using a 3 T Siemens Allegra scanner using a TR of 2 seconds. Each
fMRI dataset contains 180 time points. No motion scrubbing has been
performed, but the first four volumes were dropped in the processing
to obtain 176 time points. The ABIDE Preprocessed data have been processed
with four different pipelines, and we use the data from the CCS (connectome
computation system) pipeline here. We use the data preprocessed without
global signal regression and without bandpass filtering, as bandpass
filtering will substantially change the autoregressive structure and
we prefer to model it. Interested readers are referred to ABIDE preprocessed
for preprocessing details. As all the preprocessed data are freely
available, other researchers can reproduce our findings.

We select the ROI:s guided by \citet{Easson2019}, as their rs-fMRI
dataset also included ASD patients and healthy controls. We include
regions belonging to networks which are active during resting-state
scans for both groups, as well as there being some indication of between-group
differences in network configuration. The present analyses use ten
regions (five in each hemisphere) belonging to the Default-Mode Network
(DMN) and ten regions (also here five in each hemisphere) belonging
to the Sensory-Motor Network (SMN). More details on the selected regions
are given in Appendix D. To make graphs more readable, we refer to
the 20 brain regions by numbering them as R1-R20 instead of naming
them in the graphs (see Appendix D for a full list of region locations).
Regions R1-R10 belong to the DMN and regions R11-R20 to the SMN.

\subsection{Results on effective and functional connectivity\label{subsec:ECFCres}}

We present results on both effective (EC) and functional (FC) connectivity
for the three models in Section \ref{sec:Models}, using the data
described in Section \ref{subsec:Data}. The EC and FC results are
presented for two time lags ($L=2$) for each BVAR model, which is
the optimal number of lags for Model 1 by the widely applicable information
criterion (WAIC, \citet{Vehtari2017}), see Table \ref{table: WAIC}.
Models with two time lags for rs-fMRI data have also been suggested
previously in the literature (\citealt{Chiang2017}, \citealt{Gorrostieta2012a,Gorrostieta2013}).
\begin{table}[b]
\begin{centering}
\begin{tabular}{llrrrrrrr}
\hline 
 &  &  & Controls &  &  &  & ASDS & \tabularnewline
 &  & Model 1 & Model 2 & Model 3 & \quad{} & Model 1 & Model 2 & Model 3\tabularnewline
\cline{3-5} \cline{4-5} \cline{5-5} \cline{7-9} \cline{8-9} \cline{9-9} 
$L=1$ &  & 641306 & 646658 & 669358 &  & 645569 & 652108 & 674576\tabularnewline
$L=2$ &  & \textbf{641158} & 646460 & 669078 &  & \textbf{645288} & 651764 & 673966\tabularnewline
$L=3$ &  & 641179 & 646469 & 669109 &  & 645360 & 651769 & 673830\tabularnewline
\hline 
\end{tabular}
\par\end{centering}
\caption{WAIC information criteria for the three models with $p=20$ regions
and different lag lengths $L$.\label{table: WAIC}}
\end{table}
 In addition, note that Model 1 is superior to Model 2 and 3 for each
lag and group with substantially lower values of WAIC. Hence, the
results clearly suggest that the heterogeneous subject-specific covariance
matrices in Model 1 are indeed needed for modeling this data. .

\begin{figure}[h!]
\begin{centering}
\includegraphics[scale=0.6]{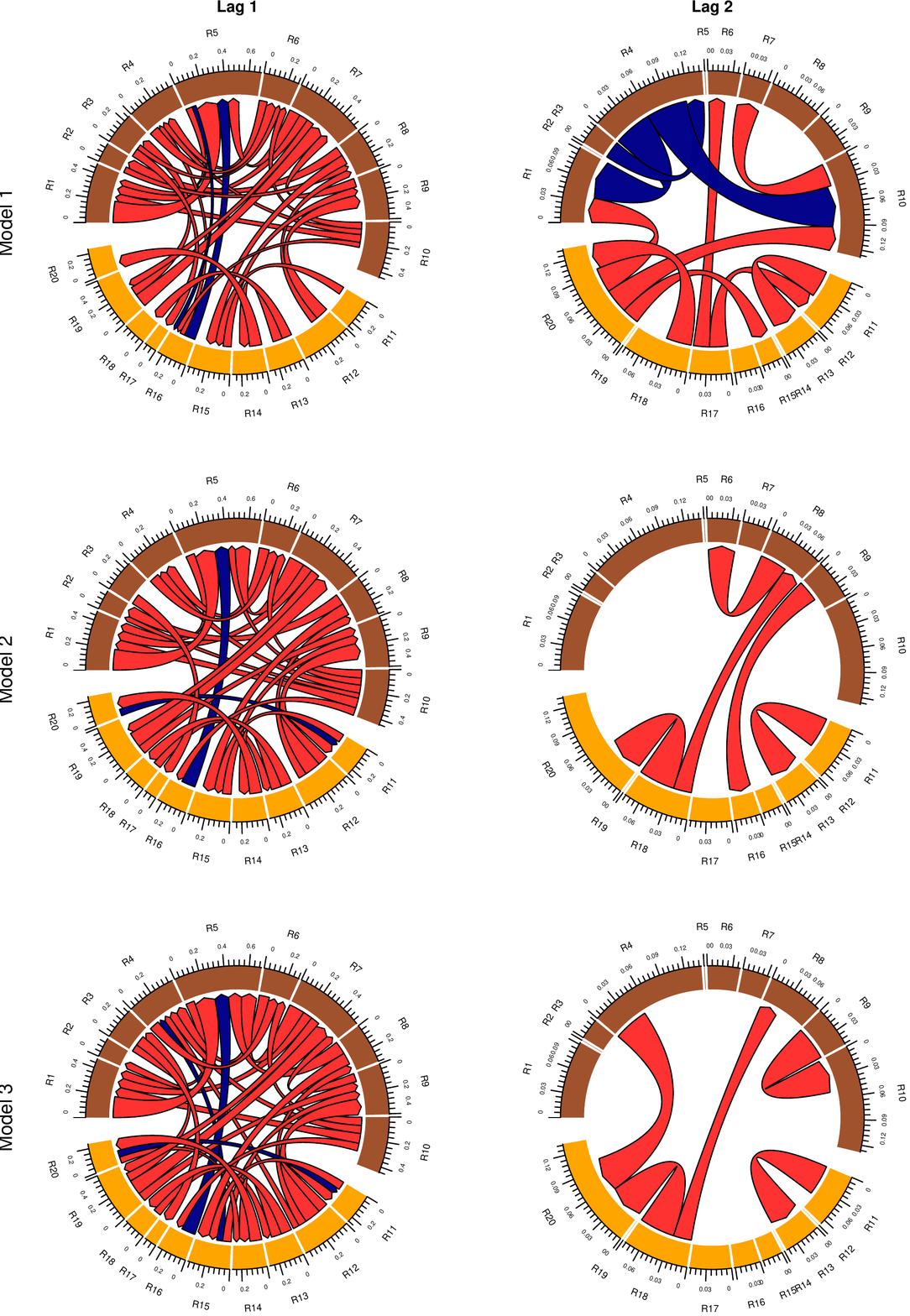}
\par\end{centering}
\caption{Control group EC: Effective connections, measured by AR coefficients,
for the first (left column) and second (right column) time lag for
each of our three models (rows one, two and three, respectively).
The regions in the DMN are R1-R10 (brown), regions R11-R20 (orange)
are the SMN regions. Thresholds for the first time lag were set to
the following: connections for which all posterior draws lie above
or below zero were kept. For the second time lag we retained connections
whose 95\% credible intervals do not include zero. The arc length
for each region corresponds to the sum of the absolute value of coefficients
for the model with the highest sum after thresholding. Direction of
connections are indicated with an arrowhead towards the \textquotedblleft receiving\textquotedblright{}
region, thickness of the arrow indicate connection strength (coefficient
size), and the color of the arrow indicate a positive (red) or negative
(blue) coefficient. \label{Fig1}}
\end{figure}

\begin{figure}[h!]
\begin{centering}
\includegraphics[scale=0.6]{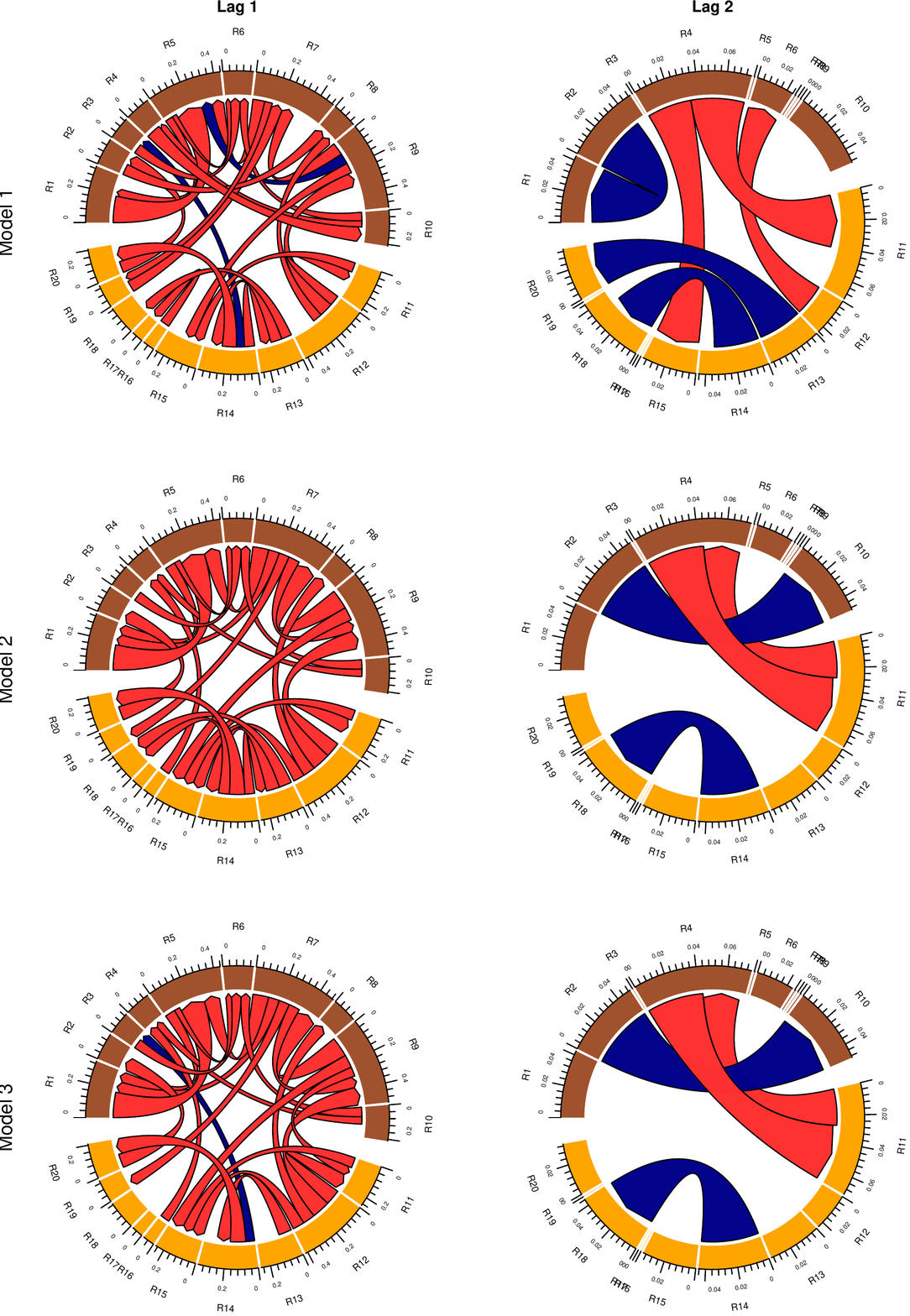}
\par\end{centering}
\caption{ASD group EC: Effective connections, measured by AR coefficients,
the first (left column) and second (right column) time lag for each
of our three models (rows one, two and three, respectively). The regions
in the DMN are R1-R10 (brown), regions R11-R20 (orange) are the SMN
regions. Thresholds for the first time lag were set to the following:
connections for which all posterior draws lie above or below zero
were kept. For the second time lag we retained connections whose 95\%
credible intervals do not include zero. The arc length for each region
corresponds to the sum of the absolute value of coefficients for the
model with the highest sum after thresholding. Direction of connections
are indicated with an arrowhead towards the \textquotedblleft receiving\textquotedblright{}
region, thickness of the arrow indicate connection strength (coefficient
size), and the color of the arrow indicate a positive (red) or negative
(blue) coefficient. \label{Fig2}}
\end{figure}

Results on EC corresponds to posterior inference on the AR-coefficients
in our Bayesian VAR models. For illustration purposes, we apply thresholds
on the posterior distribution of the AR-coefficients, where the threshold
is applied on both size of the coefficient and whether a 90 or 95
\% credible interval includes zero or not. We show results on EC for
each of the ASD and control groups separately, as well as the group
differences in EC at different time lags. Results for the separate
groups are seen in Figures \ref{Fig1} (control group) and \ref{Fig2}
(ASD group).
Each subgraph gives a visual network description of the
thresholded directed connections between regions. The general pattern
across groups and models is that there are considerably more connections
at one time lag than at two time lags, despite having a stricter threshold
at one time lag. There are more connections within a network than
between networks, as expected for the definition of the network, and
most of the coefficients are positive. For a given time lag, there
are some differences between the models. Model 2 and 3 yield more
connections than Model 1 for lag 1, while Model 1 yields some additional,
mostly negative, connections compared to Model 2 and 3.

\begin{figure}[h!]
\begin{centering}
\includegraphics[scale=0.6]{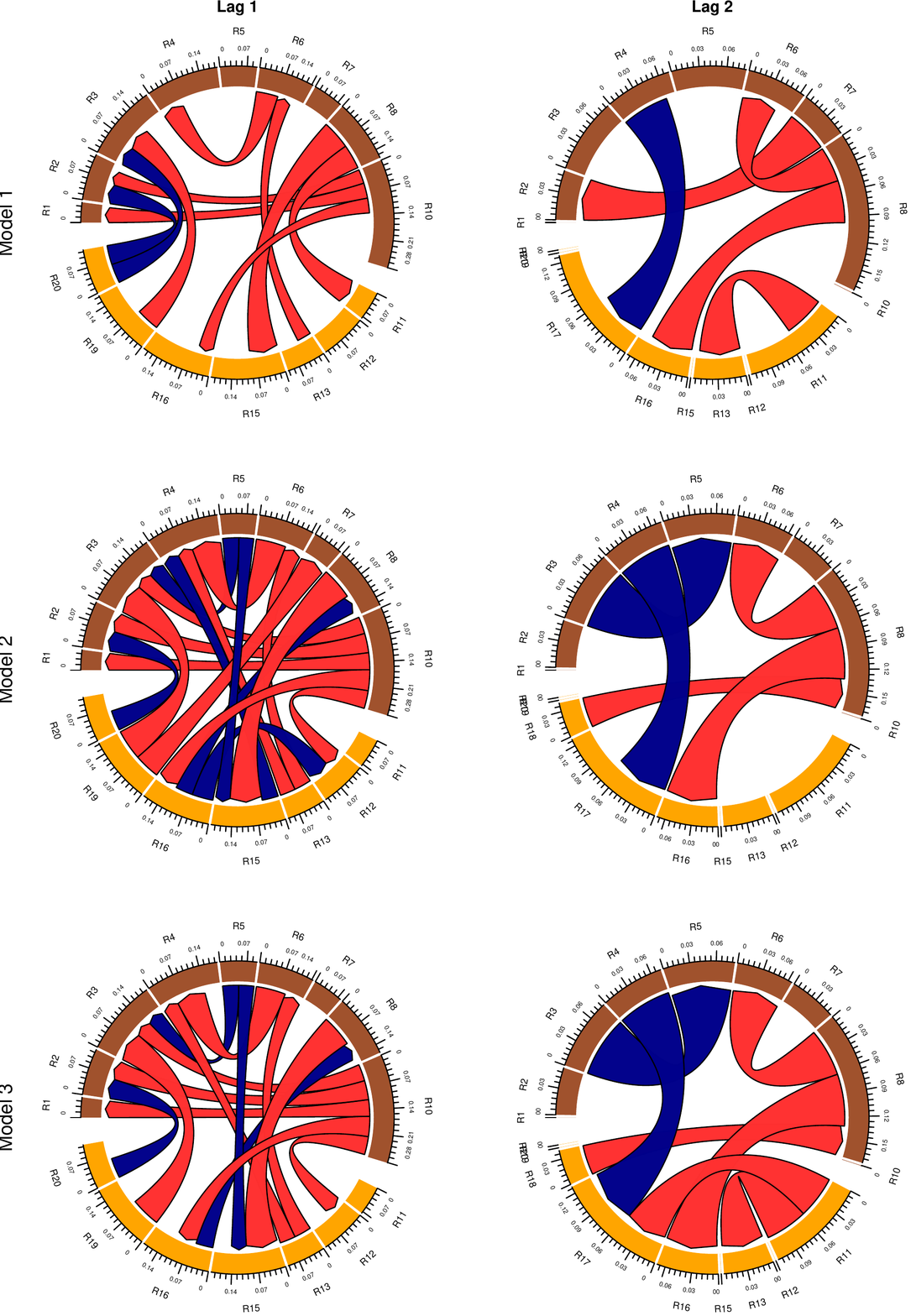}
\par\end{centering}
\caption{Difference in EC, Control group EC -- ASD group EC: Effective connection
difference, measured by AR coefficient difference, for one (left column)
and two (right column) time lags, for each of our three models (rows
one, two and three, respectively). The regions in the DMN are R1-R10
(brown), regions R11-R20 (orange) are the SMN regions. Thresholds
for the first time lag was set to the following: connections whose
95\% credible intervals do not include zero, were kept. For the second
time lag we retained connections whose 95\% credible intervals do
not include zero. The arc length for each region corresponds to the
sum of the absolute value of coefficient difference for the model
with the highest sum after thresholding. Direction of connections
are indicated with an arrowhead towards the \textquotedblleft receiving\textquotedblright{}
region, thickness of the arrow indicate size of coefficient difference,
and the color of the arrow indicate a positive (red) or negative (blue)
coefficient difference. \label{Fig3}}
\end{figure}

Figure \ref{Fig3} illustrates differences in EC between the groups
as the difference in corresponding AR-coefficients.
Results from Models 2 and 3 indicate substantially more group differences than Model 1
for both time lags, while for a given model the number of differences
is greater for lag 1. In the figure, the differences for the two lags
look comparable, but note that we use a more leniant threshold for
lag 2 for illustrative purposes (otherwise there would have been only
one connection for the differences of lag 2). In general, most of
the differences are within-network, as for the group-specific connections.
There are also some discrepancies in EC between Models 2 and 3, but
much less than the corresponding discrepancies between any of these
models and Model 1.

\begin{figure}[h!]
\begin{centering}
\includegraphics[scale=0.4]{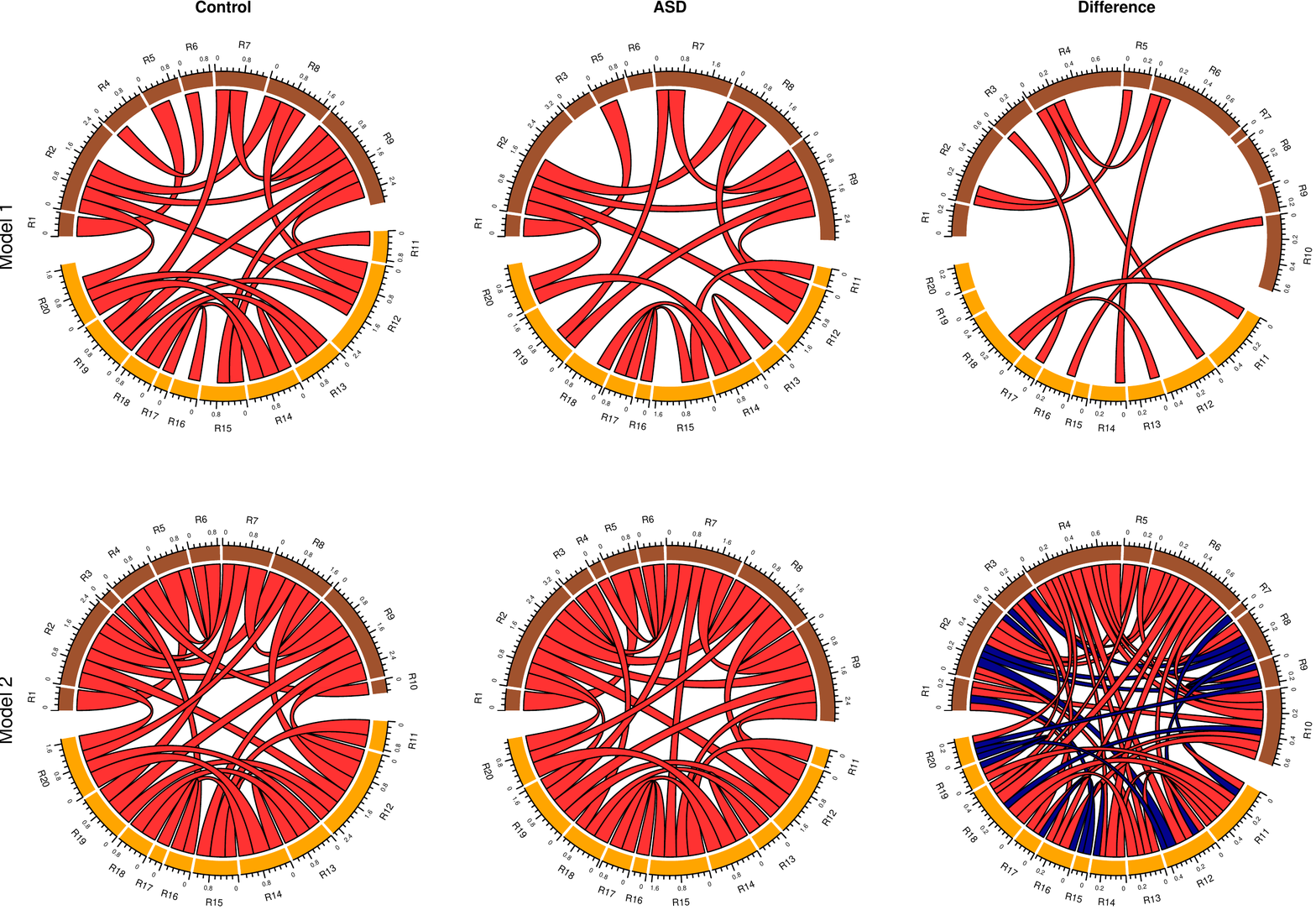}
\par\end{centering}
\caption{FC for both groups and the group difference in FC, measured by the
estimated FC correlation matrix between regions and the FC correlation
matrix difference between regions (columns) for Models 1 and 2 (rows).
The regions in the DMN are R1-R10 (brown), regions R11-R20 (orange)
are the SMN regions. The following thresholding was used in the plot:
for the separate groups, the absolute posterior mean correlation was
thresholded at 0.35, and all posterior draws lie above or below zero.
When considering the group difference, the absolute posterior mean
difference had to exceed 0.05 and the 95\% credible intervals exclude
zero. The arc length for each region corresponds to the sum of the
absolute value of the correlations - or the correlation differences
- for the model with the highest sum after thresholding. Thickness
of the line indicate size of correlation/correlation difference, and
the color of the arrow indicate a positive (red) or negative (blue)
coefficient/coefficient difference. \label{Fig4}}
\end{figure}

Results for FC are shown in Figure \ref{Fig4}.
We only compare Models 1 and 2, since Model 3 has a diagonal covariance matrix and therefore
does not allow estimation of FC. We follow the same procedure as for
EC by considering FC for each group separately, as well as the difference
in FC between groups. The figures were constructed in a similar manner
as for the figures for EC, but FC is undirected such that connection
lines between regions are undirected. It is clear that the number
of functional connections is considerably different between the models
in the figure, where Model 2 yields many more functional connections
than Model 1. Hence, it is important to account for heterogeneous
subject-specific covariance matrices in Model 1 compared to a common
covariance matrix in Model 2 in order to obtain accurate inference
on functional connectivity. The overestimation of functional connections
in Model 2 also implies an overestimation of group differences, resulting
in many spurious functional connections between the groups from Model
2.

\begin{figure}[h!]
\begin{centering}
\includegraphics[scale=0.6]{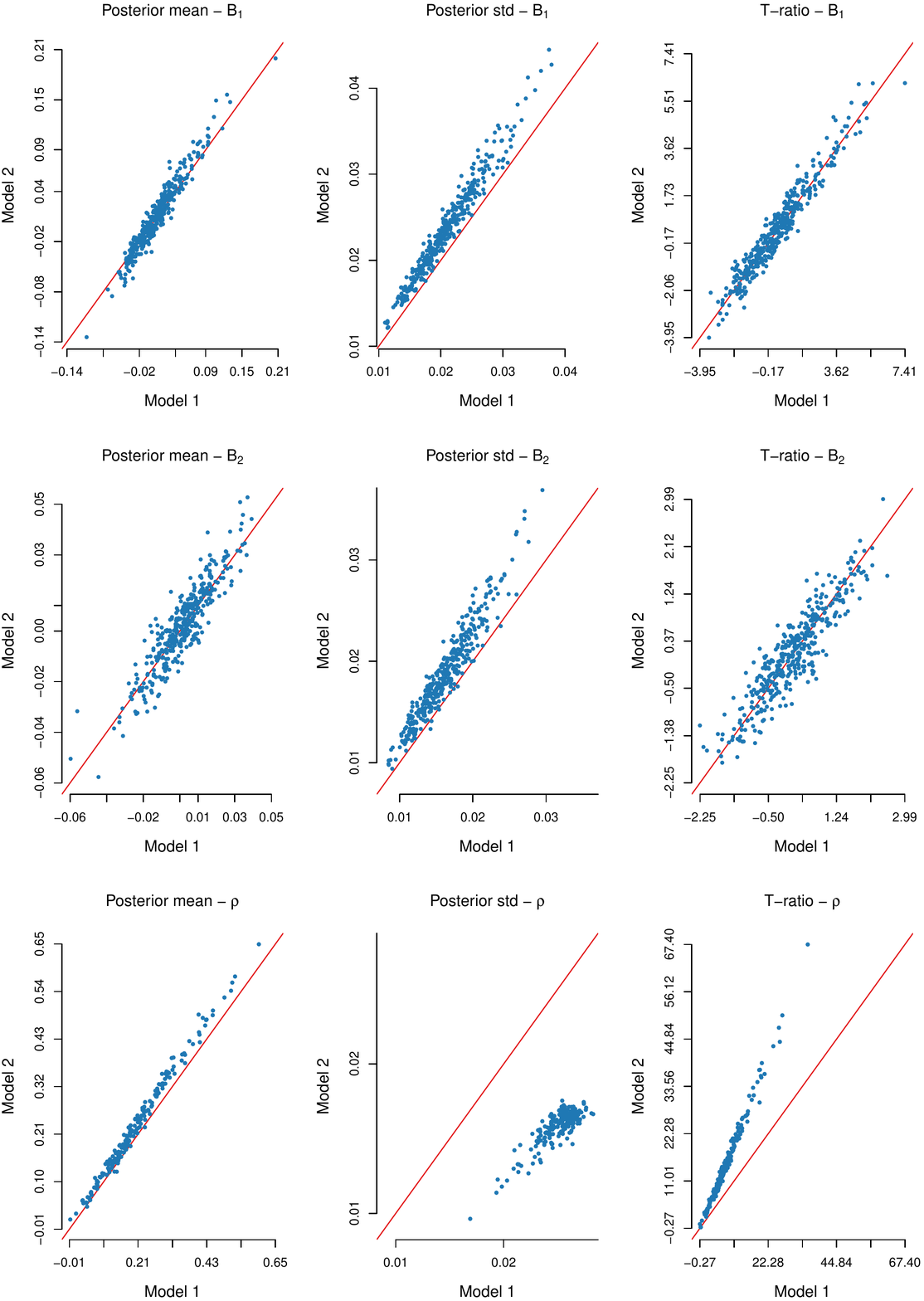}
\par\end{centering}
\caption{Scatterplots of posterior means, standard deviation and their ratio
of AR coefficients and correlations obtained using Models 1 and 2
respectively, as indicated on the axes. \label{Fig5}}
\end{figure}

We elaborate further on comparing results from the different models
by pairwise comparisons of the posterior means of the parameters.
In Figure \ref{Fig5}, AR-coefficients and the elements of the FC
correlation matrix are compared between the models.
The comparison between Model 1 and 3 was omitted as it is virtually undistinguishable
from the Model 1-Model 2 comparison. We only present this comparison
for the control group, as the results for the ASD group were very
similar. The posterior means of the AR coefficients from Models 1
and 2 are quite close (upper two panels in the left column of Figure
\ref{Fig5}). The posterior standard deviations of the respective
coefficients are slightly higher for Model 2 compared to Model 1 (upper
two panels in the middle column), but the t-ratios are quite similar
(upper two panels in the right column). Moving on to the bottom row,
the posterior means of FC correlations are fairly similar, although
Model 1 yields slightly lower posterior mean correlations compared
to Model 2. This means that the differences in FC between the Models
are mainly due to the standard deviation of the respective posterior
distributions being much smaller for Model 2, which is evident in
the t-ratios as well.

\subsection{Computing times\label{subsec:ComputationTimes}}

Table \ref{tab:ComputationTimes} presents computation times for applying
the HMC algorithm to the posterior of Model 1 for different numbers
of brain regions; the regions are randomly selected for this purpose,
since interest here is only on the computing times.
\begin{table}[t]
\begin{centering}
\begin{tabular}{ccrrr}
\hline 
 &  & $p=10$ & $p=20$ & $p=30$\tabularnewline
\cline{2-5} \cline{3-5} \cline{4-5} \cline{5-5} 
$L=1$ &  & 0.005 & 0.094 & 0.479\tabularnewline
$L=2$ &  & 0.008 & 0.385 & 9.242\tabularnewline
$L=3$ &  & 0.018 & 1.313 & 48.805\tabularnewline
$L=4$ &  & 0.048 & 5.288 & N/A\tabularnewline
\hline 
\end{tabular}
\par\end{centering}
\caption{Computation times (hours) for Model $1$ with different lag lengths
$L$ and number of regions $p$. The data consists of $S=20$ randomly
selected healthy controls and the length of each time series within
each region is $T=176$. The computer analyses were run using two
parallel MCMC chains with $200$ warm-up and $500$ sampling draws
for each chain on two CPUs with a 2.7GHz processor. N/A (not applicable)
means that the HMC algorithm took too long time to converge and was
therefore interrupted.\label{tab:ComputationTimes}}
\end{table}
Our approach works fine for around 20-30 brain regions compared to previous VAR modeling
of 5-6 regions, e.g. \citet{Chiang2017,Gorrostieta2012a,Gorrostieta2013},
but becomes very computational demanding for more than $p=30$ regions
and at least two time lags.

\section{Conclusions}

We propose a novel Bayesian VAR hierarchical model for brain connectivity
that accounts for autoregressive dependencies within and between brain
regions, and apply it to an existing, openly available rs-fMRI data
set. Compared to existing Bayesian VAR hierarchical models for this
purpose, we incorporate more flexible, subject-specific, covariance
modeling that estimates both effective and functional connectivity
simultaneously. By using the information criteria WAIC, we show that
our proposed model is superior to special cases of our model with
a common covariance matrix for all subjects. Similar simplified, diagonal,
covariance matrices has been used previously for effective connectivity
by \citet{Chiang2017} and \citet{Gorrostieta2012a,Gorrostieta2013}.
We are also able to handle 20-30 brain regions compared to previous
VAR modeling in the literature of typically 5-6 regions.

Overall, our flexible model displayed the most conservative results
with respect to the number of effective and functional connections
deemed to be non-zero from common thresholds. This is especially true
for functional connectivity where the special case with a common covariance
matrix for all subjects substantially overestimates the number of
non-zero connections. We find that the standard deviations of the
corresponding posteriors to functional connectivity are much lower
for the special model case, which implies that between-subject variation
is underestimated in such models.

We suggest some future extensions of our work. Flexible VAR modeling
for effective and functional connectivity is not yet ready for large-scale
brain connectivity, which typically involves hundreds of brain regions.
Our derived, analytical result for the posterior inference of the
model with a common covariance matrix for all subjects can be directly
applied, but with the obvious drawback of biased inference for effective
and, especially, functional connectivity. Another possibility can
be to extend our modeling to handle longitudinal data, e.g. repeated
rs-fMRI scans over time.

\begin{appendix}
\section{Posterior distribution of $\left(B,\Sigma\right)$ for
Model 2 with common covariance matrix $\Sigma$ \label{Appendix A}}
Let $X_{s,t}=\left(Y_{s,t-1},\dots,Y_{s,t-L}\right)$ be the set of
covariates in the BVAR model. The likelihood function of $\left(B_{s},\Sigma\right)$
for each subject $s$ is given by
\[
p\left(Y_{s}|X_{s},B_{s},\Sigma\right)=\prod_{t=L+1}^{T}\left|2\pi\Sigma\right|^{-1/2}\exp\left(-\frac{1}{2}\left(Y_{s,t}-B_{s}^{T}X_{s,t}\right)^{T}\Sigma^{-1}\left(Y_{s,t}-B_{s}^{T}X_{s,t}\right)\right)
\]
\[
=\left|2\pi\Sigma\right|^{-n/2}\exp\left(-\frac{1}{2}\textrm{tr}\Sigma^{-1}\left(Y_{s}-X_{s}B_{s}\right)^{T}\left(Y_{s}-X_{s}B_{s}\right)\right).
\]
Completing the squares of $B_{s}$, the likelihood function of $\left(B_{s},\Sigma\right)$
can be written as 
\[
p\left(Y_{s}|X_{s},B_{s},\Sigma\right)=\left|2\pi\Sigma\right|^{-n/2}\exp\left(-\frac{1}{2}\textrm{tr}\Sigma^{-1}V_{s}\right)\exp\left(-\frac{1}{2}\textrm{tr}\left(B_{s}-\hat{B_{s}}\right)^{T}X_{s}^{T}X_{s}\left(B_{s}-\hat{B_{s}}\right)\Sigma^{-1}\right),
\]
where $n=T-L$, $V_{s}=\left(Y_{s}-X_{s}\hat{B_{s}}\right)^{T}\left(Y_{s}-X_{s}\hat{B_{s}}\right)$
and $\hat{B_{s}}=\left(X_{s}^{T}X_{s}\right)^{-1}X_{s}^{T}Y_{s}.$
Then, using the identity 
\[
\textrm{tr}\left(A_{1}^{T}A_{2}A_{3}A_{4}^{T}\right)=\left(\textrm{vec}A_{1}\right)^{T}\left(A_{4}\otimes A_{2}\right)\left(\textrm{vec}A_{3}\right)
\]
with $A_{1}=A_{3}=B_{s}-\hat{B_{s}}$, $A_{2}=X_{s}^{T}X_{s}$, $A_{4}=\Sigma^{-1},$
the likelihood function of $\left(B_{s},\Sigma\right)$ becomes 
\[
p\left(Y_{s}|X_{s},B_{s},\Sigma\right)=\left|2\pi\Sigma\right|^{-n/2}\exp\left(-\frac{1}{2}\textrm{tr}\Sigma^{-1}V_{s}\right)\exp\left(-\frac{1}{2}\left(B_{s}-\hat{B_{s}}\right)^{T}\left(\Sigma^{-1}\otimes X_{s}^{T}X_{s}\right)\left(B_{s}-\hat{B_{s}}\right)\right).
\]
The marginal likelihood function of $\left(B,\Sigma\right)$ for all
subjects becomes 
\[
\prod_{s=1}^{S}p\left(Y_{s}|B,\Sigma,X_{s}\right)=\prod_{s=1}^{S}\int p\left(Y_{s}|X_{s},B_{s},\Sigma\right)p\left(B_{s}|\Sigma\right)dB_{s}
\]
\[
=c_{0}\prod_{s=1}^{S}c_{\kappa}\left|\Sigma\right|^{-n/2}\exp\left(-\frac{1}{2}\textrm{tr}\Sigma^{-1}\left[\tilde{V_{s}}+\left(\tilde{B_{s}}-B\right)^{T}P_{s}\left(\tilde{B_{s}}-B\right)\right]\right),
\]
where $c_{\kappa}=\prod_{s=1}^{S}c_{\kappa_{s}}=\prod_{s=1}^{S}\left(\left|P_{s}\right|^{p/2}\left|P_{s}+X_{s}^{T}X_{s}\right|^{-p/2}\right),$
$\tilde{V_{s}}=\left(Y_{s}-X_{s}\tilde{B}_{s}\right)^{T}\left(Y_{s}-X_{s}\tilde{B}_{s}\right),$
$\tilde{B_{s}}=\left(P_{s}+X_{s}^{T}X_{s}\right)^{-1}\left(X_{s}^{T}Y_{s}+P_{s}B\right),$
and $c_{0}$ is a constant that does not depend on $B$ and $\Sigma$.
Rewriting this marginal likelihood on a quadratic form of $B$ and
then multiplying the marginal likelihood with the prior distribution
of $\left(B,\Sigma\right)$, the posterior distribution of $\left(B,\Sigma\right)$
becomes
\[
p\left(B,\Sigma|Y_{s},X_{s}\right)=c_{0}c_{\kappa}\left|P_{0}\right|^{p/2}\left|\Sigma\right|^{-\left(Sn+\nu_{0}+p+1\right)/2}\exp\left(-\frac{1}{2}\textrm{tr}\Sigma^{-1}\left(\nu_{0}\Psi_{0}+\left(B-B_{0}\right)^{T}P_{0}\left(B-B_{0}\right)\right)\right)
\]
\[
\times\prod_{s=1}^{S}\exp\left(-\frac{1}{2}\textrm{tr}\Sigma^{-1}\left[R_{s}+\left(B-E_{s}\right)^{T}Q_{s}^{-1}\left(B-E_{s}\right)\right]\right),
\]
where $R_{s}=\left(Y_{s}-X_{s}K_{1s}X_{s}^{T}Y_{s}\right)^{T}\left(Y_{s}-X_{s}K_{1s}X_{s}^{T}Y_{s}\right)+Y_{s}^{T}X_{s}K_{1s}P_{s}K_{1s}X_{s}^{T}Y_{s}-E_{s}^{T}Q_{s}^{-1}E_{s}$,
\\
$K_{1s}=\left(P_{s}+X_{s}^{T}X_{s}\right)^{-1},$ $Q_{s}=\left(P_{s}K_{1s}X_{s}^{T}X_{s}K_{1s}P_{s}+\left(I-K_{1s}P_{s}\right)^{T}P_{s}\left(I-K_{1s}P_{s}\right)\right)^{-1},$
\\
 $E_{s}=Q_{s}\left(P_{s}K_{1s}X_{s}^{T}\left(Y_{s}-X_{s}K_{1s}X_{s}^{T}Y_{s}\right)+\left(I-K_{1s}P_{s}\right)^{T}P_{s}K_{1s}X_{s}^{T}Y_{s}\right),$
and $I$ is the identity matrix.\\
Rewriting on a quadratic form of $B$, the posterior distribution
is finally given by 
\[
p\left(B,\Sigma|Y_{s},X_{s}\right)=c_{0}c_{\kappa}\left|P_{0}\right|^{p/2}\left|\Sigma\right|^{-\left(Sn+\nu_{0}+p+1\right)/2}\exp\left(-\frac{1}{2}\textrm{tr}\Psi_{n}\Sigma^{-1}\right)
\]
\[
\times\exp\left(-\frac{1}{2}\textrm{tr}\Sigma^{-1}\left(B-\tilde{B}\right)^{T}\tilde{P}\left(B-\tilde{B}\right)\right),
\]
where $\Psi_{n}=\nu_{0}\Psi_{0}+\sum_{s=1}^{S}\left(R_{s}+E_{s}^{T}Q_{s}^{-1}E_{s}\right)+B_{0}^{T}P_{0}B_{0}-\tilde{B}^{T}\tilde{P}\tilde{B},$
$\tilde{P}=P_{0}+\sum_{s=1}^{S}Q_{s}^{-1}$ and $\tilde{B}=\tilde{P}^{-1}\left(P_{0}B_{0}+\sum_{s=1}^{S}Q_{s}^{-1}E_{s}\right)$.
\section{Marginal likelihood of $\left(B,\Sigma,\nu\right)$ for
Model 1 with subject-specific covariance matrix $\Sigma_{s}$ \label{Appendix B}}
Replacing $\Sigma$ with $\Sigma_{s}$ in Equation \eqref{eq:Likelihood}
gives the likelihood function of $\left(B_{s},\Sigma_{s}\right)$
for each subject $s$ as 
\[
p\left(Y_{s}|X_{s},B_{s},\Sigma_{s}\right)=\left|2\pi\Sigma_{s}\right|^{-n/2}\exp\left(-\frac{1}{2}\textrm{tr}\Sigma_{s}^{-1}\left(Y_{s}-X_{s}B_{s}\right)^{T}\left(Y_{s}-X_{s}B_{s}\right)\right).
\]
The marginal likelihood function of $\left(B,\Sigma\right)$ for all
subjects becomes 
\[
p\left(Y|B,\Sigma\right)=\prod_{s=1}^{S}p\left(Y_{s}|B,\Sigma,X_{s}\right)=\prod_{s=1}^{S}\int\int p\left(Y_{s}|B_{s},\Sigma_{s},X_{s}\right)p\left(B_{s},\Sigma_{s}\right)dB_{s}d\Sigma_{s}
\]
\[
=c_{1}\left|\nu\Sigma\right|^{\nu/2}\prod_{s=1}^{S}\left|\nu\Sigma+\tilde{V_{s}}+\left(\tilde{B_{s}}-B\right)^{T}P_{s}\left(\tilde{B_{s}}-B\right)\right|^{-\left(n+\nu\right)/2},
\]
where $c_{1}$ is a constant that does not depend on $B$ and $\Sigma$.
Rewriting on a quadratic form of $B$ in the determinant, the marginal
likelihood of $\left(B,\Sigma\right)$ can be expressed as

\[
p\left(Y|B,\Sigma\right)=c\left|\nu\Sigma\right|^{\nu/2}\prod_{s=1}^{S}\left|\nu\Sigma+R_{s}+\left(B-E_{s}\right)^{T}Q_{s}^{-1}\left(B-E_{s}\right)\right|.
\]
\section{Stan modeling code for Model 1 \label{StanCode}}
\texttt{\textbf{\textcolor{blue}{\scriptsize{}data}}}\texttt{\scriptsize{}
\{}{\scriptsize\par}

\texttt{\scriptsize{}int<lower=0> p; }\texttt{\textcolor{olive}{\scriptsize{}//
number of brain regions}}{\scriptsize\par}

\texttt{\scriptsize{}int<lower=0> q; }\texttt{\textcolor{olive}{\scriptsize{}//
number of covariates L{*}p}}{\scriptsize\par}

\texttt{\scriptsize{}int<lower=0> S; }\texttt{\textcolor{olive}{\scriptsize{}//
number of subjects}}{\scriptsize\par}

\texttt{\scriptsize{}int<lower=0> qp; }\texttt{\textcolor{olive}{\scriptsize{}//
number of VAR coefficients, qp = q{*}p}}{\scriptsize\par}

\texttt{\scriptsize{}int<lower=0> T; }\texttt{\textcolor{olive}{\scriptsize{}//
number of time points}}{\scriptsize\par}

\texttt{\scriptsize{}matrix {[}p,p{]} R\_s{[}S{]}; }\texttt{\textcolor{olive}{\scriptsize{}//
array with matrices R\_s for all subjects}}{\scriptsize\par}

\texttt{\scriptsize{}matrix {[}q,p{]} E\_s{[}S{]}; }\texttt{\textcolor{olive}{\scriptsize{}//
array with matrices E\_s for all subjects}}{\scriptsize\par}

\texttt{\scriptsize{}matrix {[}q,q{]} Q\_s\_inv{[}S{]}; }\texttt{\textcolor{olive}{\scriptsize{}//
array with matrices Q\_s\_inv for all subjects}}{\scriptsize\par}

\texttt{\textcolor{olive}{\scriptsize{}// Prior settings}}{\scriptsize\par}

\texttt{\scriptsize{}vector{[}qp{]} B\_0\_spec; }\texttt{\textcolor{olive}{\scriptsize{}//
prior Mean}}{\scriptsize\par}

\texttt{\scriptsize{}cov\_matrix{[}q{]} Chol\_Cov\_B; }\texttt{\textcolor{olive}{\scriptsize{}//
cholesky decomposition of the covariance matrix for B}}{\scriptsize\par}

\texttt{\scriptsize{}int nu\_0; }\texttt{\textcolor{olive}{\scriptsize{}//
degrees of freedom in the prior for Sigma}}{\scriptsize\par}

\texttt{\scriptsize{}cov\_matrix{[}p{]} Psi\_0; }\texttt{\textcolor{olive}{\scriptsize{}//
scale matrix in the prior for Sigma}}{\scriptsize\par}

\texttt{\scriptsize{}cov\_matrix{[}qp{]} I\_Mat; }\texttt{\textcolor{olive}{\scriptsize{}//
identity matrix}}{\scriptsize\par}

\texttt{\scriptsize{}\}}{\scriptsize\par}

\texttt{\textbf{\textcolor{blue}{\scriptsize{}parameters}}}\texttt{\scriptsize{}
\{}{\scriptsize\par}

\texttt{\scriptsize{}cov\_matrix{[}p{]} Sigma; }\texttt{\textcolor{olive}{\scriptsize{}//
covariance matrix}}{\scriptsize\par}

\texttt{\scriptsize{}matrix{[}q,p{]} B\_spec; }\texttt{\textcolor{olive}{\scriptsize{}//
matrix of VAR coefficients}}{\scriptsize\par}

\texttt{\scriptsize{}real<lower=p+2> nu; }\texttt{\textcolor{olive}{\scriptsize{}//
degrees of freedom in the prior for Sigma\_s}}{\scriptsize\par}

\texttt{\scriptsize{}\}}{\scriptsize\par}

\texttt{\textbf{\textcolor{blue}{\scriptsize{}transformed parameters}}}\texttt{\scriptsize{}
\{}{\scriptsize\par}

\texttt{\scriptsize{}matrix{[}q,p{]} B; }\texttt{\textcolor{olive}{\scriptsize{}//
matrix of VAR coefficients}}{\scriptsize\par}

\texttt{\scriptsize{}B = Chol\_Cov\_B {*} B\_spec {*} cholesky\_decompose(Sigma);}{\scriptsize\par}

\texttt{\scriptsize{}\}}{\scriptsize\par}

\texttt{\textbf{\textcolor{blue}{\scriptsize{}model}}}\texttt{\scriptsize{}
\{}{\scriptsize\par}

\texttt{\scriptsize{}real Sum\_logdet;}{\scriptsize\par}

\texttt{\scriptsize{}matrix{[}p,p{]} Part\_s;}{\scriptsize\par}

\texttt{\textcolor{olive}{\scriptsize{}// priors}}{\scriptsize\par}

\texttt{\scriptsize{}Sigma \textasciitilde{} inv\_wishart(nu\_0,Psi\_0);
}\texttt{\textcolor{olive}{\scriptsize{}// prior for the covariance
matrix Sigma}}{\scriptsize\par}

\texttt{\scriptsize{}to\_vector(B\_spec) \textasciitilde{} multi\_normal(B\_0\_spec,I\_Mat);
}\texttt{\textcolor{olive}{\scriptsize{}// special prior for parameterization}}{\scriptsize\par}

\texttt{\scriptsize{}\ Sum\_logdet = 0;}{\scriptsize\par}

\texttt{\textcolor{olive}{\scriptsize{}// log-likelihood}}{\scriptsize\par}

\texttt{\scriptsize{}\ for (s in 1:S)\{}{\scriptsize\par}

\texttt{\scriptsize{}\ \ \ Part\_s = nu{*}Sigma + R\_s{[}s{]} +
quad\_form(Q\_s\_inv{[}s{]} , B-E\_s{[}s{]});}{\scriptsize\par}

\texttt{\scriptsize{}\ \ \ Sum\_logdet = Sum\_logdet + log\_determinant(Part\_s);}{\scriptsize\par}

\texttt{\scriptsize{}\}}{\scriptsize\par}

\texttt{\scriptsize{}target += S{*}( lmgamma(p,0.5{*}(T+nu)) - lmgamma(p,0.5{*}nu)
);}{\scriptsize\par}

\texttt{\scriptsize{}target += 0.5{*}S{*}nu{*}log\_determinant(nu{*}Sigma)
- 0.5{*}(nu+T){*}Sum\_logdet;}{\scriptsize\par}

\texttt{\scriptsize{}\}}{\scriptsize\par}
\section{ROI information \label{ROI info}}
Information on our selected ROI:s in the Default-Mode Network (DMN) and Sensory-Motor Network (SMN) is given below in the following order: 
abbreviation in the manuscript, type of network the ROI is classified to, volume of the ROI, 
$(x,y,z)-$coordinates for the ROI center of mass, and AAL atlas annotation (\citealt{TzourioMazoyer2002}).\medskip

\begin{tabular}{|c|c|c|c|c|}
\hline 
Abbreviation & Network & volume mm\textasciicircum 3 & center of mass (x,y,z) & AAL annotation\tabularnewline
\hline 
\hline 
R1 & DMN & 222 & (-6.8;45.7;7.8) & Cingulum\_Ant\_L\tabularnewline
\hline 
R2 & DMN & 267 & (9.1;-35.9;47.1) & Cingulum\_Mid\_R\tabularnewline
\hline 
R3 & DMN & 213 & (6.7;42.6;6.1) & Cingulum\_Ant\_R\tabularnewline
\hline 
R4 & DMN & 247 & (0.3;16.3;32.3) & Cingulum\_Mid\_L\tabularnewline
\hline 
R5 & DMN & 249 & (-7.9;-33.1;45.5) & Cingulum\_Mid\_L\tabularnewline
\hline 
R6 & DMN & 248 & (0.0;-0.3;42.2) & Cingulum\_Mid\_L\tabularnewline
\hline 
R7 & DMN & 247 & (-14.0;-66.3;55.9) & Precuneus\_L\tabularnewline
\hline 
R8 & DMN & 214 & (-49.2;22.9;9.3) & Frontal\_Inf\_Tri\_L\tabularnewline
\hline 
R9 & DMN & 174 & (52.1;28.0;4.9) & Frontal\_Inf\_Tri\_R\tabularnewline
\hline 
R10 & DMN & 266 & (10.3;-63.5;56.2) & Precuneus\_R\tabularnewline
\hline 
R11 & SMN & 247 & (55.2;-47.5;41.9) & Parietal\_Inf\_R\tabularnewline
\hline 
R12 & SMN & 287 & (-58.9;-30.2;-2.4) & Temporal\_Mid\_L\tabularnewline
\hline 
R13 & SMN & 222 & (61.9;-21.1;-15.6) & Temporal\_Mid\_R\tabularnewline
\hline 
R14 & SMN & 204 & (-47.8;7.9;-9.3) & Insula\_R\tabularnewline
\hline 
R15 & SMN & 210 & (-39.7;-12.9;13.3) & Insula\_L\tabularnewline
\hline 
R16 & SMN & 292 & (0.4;-15.1;51.8) & Supp\_Motor\_Area\_L\tabularnewline
\hline 
R17 & SMN & 206 & (29.9;-67.5;47.3) & Parietal\_Sup\_R\tabularnewline
\hline 
R18 & SMN & 237 & (40.7;-11.3;-3.9) & Insula\_R\tabularnewline
\hline 
R19 & SMN & 234 & (-33.9;-53.8;49.5) & Parietal\_Inf\_L\tabularnewline
\hline 
R20 & SMN & 215 & (10.6;1.3;65.9) & Supp\_Motor\_Area\_R\tabularnewline
\hline 
\end{tabular}
\end{appendix}

\begin{acks}[Acknowledgments]

Anders Eklund is also affiliated with the Center for medical image science and visualization (CMIV).
\end{acks}

\begin{funding}
Anders Lundquist was supported by Riksbankens 
Jubileumsfond, Grant number P16-028:1. 
Anders Eklund was supported in part by the Center for Industrial Information Technology (CENIIT) at Linköping University.
\end{funding}

\bibliographystyle{imsart-nameyear}
\bibliography{ref}

\end{document}